\documentclass[aps,prl,twocolumn,showpacs,superscriptaddress,preprintnumbers]{revtex4-2}
\usepackage{slashed}
\usepackage[english]{babel}
\usepackage[utf8]{inputenc}
\usepackage[colorinlistoftodos, color=green!40, prependcaption]{todonotes}
\usepackage{amsthm}
\usepackage{mathtools}
\usepackage{physics}
\usepackage{xcolor}
\usepackage{graphicx}
\usepackage[left=23mm,right=13mm,top=35mm,columnsep=15pt]{geometry} 
\usepackage{adjustbox}
\usepackage{placeins}
\usepackage[T1]{fontenc}
\usepackage{lipsum}
\usepackage{csquotes}
\usepackage{diagbox}
\usepackage{float}
\usepackage{physics}

\usepackage[normalem]{ulem}

\usepackage[pdftex, pdftitle={Article}, pdfauthor={Author}]{hyperref} 
\begin{document}

\title{
SENSEI: Search for Millicharged Particles produced in the NuMI Beam
}

\author{The SENSEI Collaboration: \\ Liron Barak}

\affiliation{\normalsize\it 
 School of Physics and Astronomy, 
 Tel-Aviv University, Tel-Aviv 69978, Israel}
 
\author{Itay M. Bloch}
\affiliation{Berkeley Center for Theoretical Physics, University of California, Berkeley, CA 94720, U.S.A.}
\affiliation{Theoretical Physics Group, Lawrence Berkeley National Laboratory, Berkeley, CA 94720, U.S.A.}

\author{Ana M. Botti}
\affiliation{\normalsize\it 
Fermi National Accelerator Laboratory, PO Box 500, Batavia IL, 60510, USA}

\author{Mariano Cababie}
\affiliation{\normalsize\it 
Department of Physics, FCEN, University of Buenos Aires and IFIBA, CONICET, Buenos Aires, Argentina}
\affiliation{\normalsize\it 
Fermi National Accelerator Laboratory, PO Box 500, Batavia IL, 60510, USA}

\author{Gustavo Cancelo}
\affiliation{\normalsize\it 
Fermi National Accelerator Laboratory, PO Box 500, Batavia IL, 60510, USA}

\author{Luke Chaplinsky}
\affiliation{\normalsize\it University of Massachusetts, Amherst Center for Fundamental Interactions and Department of Physics, Amherst, MA 01003, USA}
\affiliation{\normalsize\it 
C.N.~Yang Institute for Theoretical Physics, Stony Brook University, Stony Brook, NY 11794, USA}
\affiliation{\normalsize\it 
Department of Physics and Astronomy, Stony Brook University, Stony Brook, NY 11794, USA}

\author{Michael Crisler}
\affiliation{\normalsize\it 
Fermi National Accelerator Laboratory, PO Box 500, Batavia IL, 60510, USA}

\author{Alex Drlica-Wagner}
\affiliation{\normalsize\it 
Fermi National Accelerator Laboratory, PO Box 500, Batavia IL, 60510, USA}
\affiliation{\normalsize\it Kavli Institute for Cosmological Physics, University of Chicago, Chicago, IL 60637, USA}
\affiliation{\normalsize\it  Department of Astronomy and Astrophysics, University of Chicago, Chicago IL 60637, USA}

 \author{Rouven Essig}
\affiliation{\normalsize\it 
C.N.~Yang Institute for Theoretical Physics, Stony Brook University, Stony Brook, NY 11794, USA}

 \author{Juan Estrada}
\affiliation{\normalsize\it 
Fermi National Accelerator Laboratory, PO Box 500, Batavia IL, 60510, USA}

\author{Erez Etzion}
\affiliation{\normalsize\it 
 School of Physics and Astronomy, 
 Tel-Aviv University, Tel-Aviv 69978, Israel}

\author{Guillermo Fernandez Moroni}
\affiliation{\normalsize\it 
Fermi National Accelerator Laboratory, PO Box 500, Batavia IL, 60510, USA}

\author{Stephen E. Holland}
\affiliation{\normalsize\it 
Lawrence Berkeley National Laboratory, One Cyclotron Road, Berkeley, California 94720, USA}

\author{Yaron Korn}
\affiliation{\normalsize\it 
 School of Physics and Astronomy, 
 Tel-Aviv University, Tel-Aviv 69978, Israel}

\author{Sravan Munagavalasa}
\affiliation{\normalsize\it Kavli Institute for Cosmological Physics, University of Chicago, Chicago, IL 60637, USA}
\affiliation{\normalsize\it 
The Enrico Fermi Institute, The University of Chicago, Chicago, Illinois 60637, USA}
\affiliation{\normalsize\it 
C.N.~Yang Institute for Theoretical Physics, Stony Brook University, Stony Brook, NY 11794, USA}
\affiliation{\normalsize\it 
Department of Physics and Astronomy, Stony Brook University, Stony Brook, NY 11794, USA}

 \author{Aviv Orly}
\affiliation{\normalsize\it 
 School of Physics and Astronomy, 
 Tel-Aviv University, Tel-Aviv 69978, Israel}

\author{Santiago E. Perez}
\email{santiep.137@gmail.com}
\affiliation{\normalsize\it 
Department of Physics, FCEN, University of Buenos Aires and IFIBA, CONICET, Buenos Aires, Argentina}
\affiliation{\normalsize\it 
Fermi National Accelerator Laboratory, PO Box 500, Batavia IL, 60510, USA}

\author{Dario Rodrigues}
\affiliation{\normalsize\it 
Department of Physics, FCEN, University of Buenos Aires and IFIBA, CONICET, Buenos Aires, Argentina}
\affiliation{\normalsize\it 
Fermi National Accelerator Laboratory, PO Box 500, Batavia IL, 60510, USA}

\author{Nathan A. Saffold}
\affiliation{\normalsize\it 
Fermi National Accelerator Laboratory, PO Box 500, Batavia IL, 60510, USA}

\author{Silvia Scorza}
\affiliation{\normalsize\it 
Univ. Grenoble Alpes, CNRS, Grenoble INP, LPSC-IN2P3, Grenoble, 38000, France}

\author{Aman Singal}
\affiliation{\normalsize\it 
C.N.~Yang Institute for Theoretical Physics, Stony Brook University, Stony Brook, NY 11794, USA}
\affiliation{\normalsize\it 
Department of Physics and Astronomy, Stony Brook University, Stony Brook, NY 11794, USA}
 
\author{Miguel Sofo Haro}
\affiliation{\normalsize\it 
Fermi National Accelerator Laboratory, PO Box 500, Batavia IL, 60510, USA}
\affiliation{Centro At\'omico Bariloche, CNEA/CONICET/IB, Bariloche, Argentina}

\author{Leandro Stefanazzi}
\affiliation{\normalsize\it 
Fermi National Accelerator Laboratory, PO Box 500, Batavia IL, 60510, USA}

 \author{Kelly Stifter}
\affiliation{\normalsize\it 
Fermi National Accelerator Laboratory, PO Box 500, Batavia IL, 60510, USA}

\author{Javier Tiffenberg}
\affiliation{\normalsize\it 
Fermi National Accelerator Laboratory, PO Box 500, Batavia IL, 60510, USA}

\author{Sho Uemura}
\affiliation{\normalsize\it 
Fermi National Accelerator Laboratory, PO Box 500, Batavia IL, 60510, USA}

\author{Tomer Volansky}
\affiliation{\normalsize\it 
 School of Physics and Astronomy,   
 Tel-Aviv University, Tel-Aviv 69978, Israel}

\author{Tien-Tien Yu}
\affiliation{\normalsize\it 
Department of Physics and Institute for Fundamental Science, University of Oregon, Eugene, Oregon 97403, USA}

\author{\\In collaboration with: \\ Roni Harnik}
\affiliation{\normalsize\it 
Fermi National Accelerator Laboratory, PO Box 500, Batavia IL, 60510, USA}

\author{Zhen Liu}
\affiliation{\normalsize\it 
School of Physics and Astronomy, University of Minnesota,
Minneapolis, MN 55455, USA}

\author{Ryan Plestid}
\affiliation{\normalsize\it Walter
 Burke Institute for Theoretical Physics, California Institute of Technology, Pasadena, CA 91125}

\date{\today} % Leave empty to omit a date

\preprint{CALT-TH-2023-011, YITP-SB-2023-07, FERMILAB-PUB-23-222-PPD}

\begin{abstract}
\noindent 
 Millicharged particles appear in several extensions of the Standard Model, but have not yet been detected. These hypothetical particles could be produced by an intense proton beam striking a fixed target. We use data collected in 2020 by the SENSEI experiment in the MINOS cavern at the Fermi National Accelerator Laboratory to search for ultra-relativistic millicharged particles produced in collisions of protons in the NuMI  beam with a fixed graphite target. 
The absence of any ionization events with 3 to 6 electrons in the SENSEI data allow us to place world-leading constraints on millicharged particles for masses between 30 MeV to 380 MeV. This work also demonstrates the potential of utilizing low-threshold detectors to investigate new particles in beam-dump experiments, and motivates a future experiment designed specifically for this purpose. 
\end{abstract}


\maketitle

\section{Status of Millicharged Particle Searches} \label{sec:intro}

\noindent Fractionally charged particles are a well-known feature of the Standard Model (SM): quarks and antiquarks have an electric charge $\pm 1/3$ or $\pm 2/3$ that of the electron. However, there is no particle in the SM with a charge number less than 1/3. Millicharged particle (mCP) models are extensions of the SM where a new particle is introduced with a very small electric charge. This can be achieved by adding a U(1) symmetric Lagrangian term charged under the Standard Model hypercharge, 
\begin{equation}
    \mathcal{L}_{\rm mCP}=i\Bar{\chi}(\slashed\partial-i\varepsilon e \slashed B+ M_{mCP})\chi \,,
\end{equation}

\noindent represents $\chi$ the mCP field, $\slashed B$ is the SM electroweak vector boson and $\varepsilon$ is the millicharge. These mCPs can also appear, for example, in models where a dark photon has a small kinetic mixing with the SM photon, resulting in a very small charge for the dark sector particles, in which case a different vector boson $\slashed B'$ has to be considered ~\cite{mCPdarkPhoton}.  mCPs have also been suggested as dark matter (DM) candidates~\cite{Emken_2019}, and extensions of the SM with mCPs have been considered to describe several experimental results~\cite{Foot:2007cq,gies2006polarized, Wallemacq:2014sta,farzan2019dark,Khan:2020vaf,Farzan:2020dds}. For these reasons, mCPs in the MeV to GeV mass range are interesting targets for experiments. 

mCPs could be produced in high energy collisions at particle accelerators. Some recent accelerator experiments which looked for, or had searches recast for, mCPs include  milliQ ~\cite{Prinz:1998ua}, milliQan~\cite{Ball:2016zrp,Ball:2020dnx}, LSND~\cite{Magill:2018tbb,Auerbach:2001wg},
 MiniBooNE~\cite{Magill:2018tbb,Aguilar-Arevalo:2018wea} and ArgoNeuT~\cite{Harnik:2019zee,Acciarri:2019jly}. Future experiments with sensitivity to mCPs have been proposed at accelerator facilities~\cite{Kelly:2018brz,Liang:2019zkb,Choi:2020mbk,Foroughi-Abari:2020qar}.

mCPs could also be produced from cosmic rays in the Earth's upper atmosphere~\cite{Plestid:2020kdm}. In this case, high-energy SM particles (mainly protons) reach the atmosphere and produce showers of secondary particles. mCPs produced in these collisions could reach detectors on the Earth's surface and also underground~\cite{Plestid:2020kdm,Kachelriess:2021man,mCPheavens}. This flux of mCPs could be enhanced by several orders of magnitude due to the trapping of charged particles in the galactic magnetic field~\cite{Harnik2021}. A flux of mCPs at detectors could also be generated from cosmic-ray upscattering of a millicharged component of DM~\cite{Harnik2021}.

In this letter, we present the results of a mCP search utilizing data from the SENSEI detector, which is collinear with the NuMI beamline. These results give new constraints on the mass and millicharge parameter space for particles with MeV to GeV masses. 

\section{Skipper-CCDs as a \MakeLowercase{m}CP probe} \label{sec:develop}
\noindent Charge-coupled devices (CCDs) are pixelated silicon sensors commonly used in astronomical applications.  Recently, CCDs have demonstrated their utility in particle physics experiments, and large arrays of thick, fully-depleted CCDs have been successfully deployed in a low-background environment to search for ionization signals from particle DM~\cite{DAMIC2016}. However, conventional scientific CCDs are limited by the readout noise on the measured charge in each pixel, which is typically $\sim$2e$^-$ (RMS). The Skipper-CCD substantially reduces this noise by taking multiple, non-destructive samples of the charge in each pixel. The concept was initially proposed in  1990s~\cite{1990SPIE.1242..238C,1990ASPC....8...18J}, but the performance as a single-electron counting device was only recently demonstrated~\cite{Tiffenberg:2017aac}. Skipper-CCDs have enabled a new generation of low-mass DM searches~\cite{PhysRevLett.122.161801,sensei2018,barak2022sensei,Barak:2020fql,DAMIC-M}  and coherent elastic neutrino-nucleus scattering (CE$\nu$NS) experiments~\cite{Nasteva_2021}. Future experiments are planned using this technology~\cite{oscura_2020}. The single-electron threshold capability of Skipper-CCDs makes them an ideal tool to search for mCPs with sensitivities not accessible to other technologies.
 
The SENSEI~\cite{Barak:2020fql} Collaboration performed a search in 2020 for low-mass DM using a  $\sim$2~g Skipper-CCD. The sensor used was designed by the Lawrence Berkeley National Laboratory (LBNL), and fabricated at Teledyne/DALSA using high-resistivity ($>$18k$\Omega$-cm) silicon wafers with a thickness of 675~$\mu$m.
With a total exposure of 24 days at a shallow underground laboratory at Fermi National Accelerator Laboratory (FNAL) (107~m deep), SENSEI demonstrated the lowest rates of events containing one and two electrons in silicon detectors. Using these results, SENSEI achieved world-leading sensitivity for a wide range of sub-GeV DM masses interacting via electron recoils. Here, the results from the SENSEI DM search are used to establish a new limit on the mCP millicharge and mass parameter space. To extend the sensitivity of the search, the published analysis procedure that was applied to the 3-4$e^-$ channels is also applied to 5-6$e^-$. For each channel, we calculated the effective efficiency, defined as the fraction of pixels that pass all cuts corrected by exposure and combined with a geometric efficiency for the channels with more than 3$e^-$. This geometric efficiency accounts for mCP events with more than 2$e^-$ to be spread out more than one pixel due to charge diffusion. For each new channel, we report the number of observed events. These results are shown in Table~ \ref{tab:senseiEventCount}, including the detection efficiency and the total effective exposure for all channels considered. It is worth noting that no events with with signals from 3 to 6 electron events were observed. 

\begin{figure}[t]
    \centering
    \includegraphics[width=0.35\textwidth]{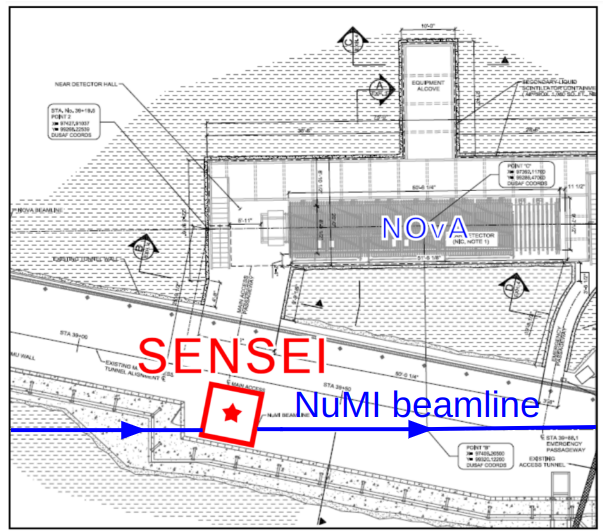}
    \caption{SENSEI detector location inside Fermilab's MINOS underground hall.}
        \label{SENSEI@MINOS_LOC}
\end{figure}

\begin{table}[t]
\begin{center}
\begin{footnotesize}
\begin{tabular}{|l|*{12}{c|}}\hline
~& \multicolumn{2}{c|}{$1e^-$} & \multicolumn{2}{c|}{$2e^-$} & \multicolumn{2}{c|}{$3e^-$} & \multicolumn{2}{c|}{$4e^-$} & \multicolumn{2}{c|}{$5e^-$} & \multicolumn{2}{c|}{$6e^-$}\\ \hline \hline
Eff. Efficiency & \multicolumn{2}{c|}{0.069} & \multicolumn{2}{c|}{0.105} & \multicolumn{2}{c|}{0.325} & \multicolumn{2}{c|}{0.327}  & \multicolumn{2}{c|}{0.331}  & \multicolumn{2}{c|}{0.338} \\ \hline
Exp.~[g-day] & \multicolumn{2}{c|}{1.38} & \multicolumn{2}{c|}{2.09} & \multicolumn{2}{c|}{9.03} & \multicolumn{2}{c|}{9.10}  & \multicolumn{2}{c|}{9.23}  & \multicolumn{2}{c|}{9.39} \\ \hline \hline
Obs.\ Events & \multicolumn{2}{c|}{1311.7} & \multicolumn{2}{c|}{5} & \multicolumn{2}{c|}{0} & \multicolumn{2}{c|}{0} &\multicolumn{2}{c|}{0} & \multicolumn{2}{c|}{0}\\ \hline
\end{tabular}
\end{footnotesize}
\caption{Performance of the SENSEI experiment for events containing 1-6~e$^-$. The efficiency here includes the effect of all selection cuts on the data (see Ref.~\cite{Barak:2020fql} for details). The bottom two rows respectively list the efficiency-corrected exposure, and the number of observed events after cuts. Compared to the results reported in~\cite{Barak:2020fql}, we add here the 5~$e^-$ and 6~$e^-$ channels.}
%\vspace{-6mm}
\label{tab:senseiEventCount}
\end{center}
\end{table}
The SENSEI detector at Fermilab is located in the MINOS underground hall. This location was selected because it provides sufficient shielding from cosmic rays that bombard Earth’s surface, and has been used in the past to test sensitive particle detector technology. The detector is located 1~km away from the target of the NuMI beam (Fig. \ref{SENSEI@MINOS_LOC}). Given that the SENSEI experiment was designed to search for DM from astrophysical sources, it is not surprising that it is not perfectly aligned with the NuMI beamline. As will be described in the following sections, the alignment and orientation of the silicon detector is not relevant because the mCPs of interest have a low probability of interaction and because the whole detector is contained in the solid angle subtended by mCPs generated from the beam.

\section{mCP production in the NuMI beamline and acceptance by the SENSEI detector}

\begin{figure}[t]
    \centering
    \includegraphics[width=0.4\textwidth]{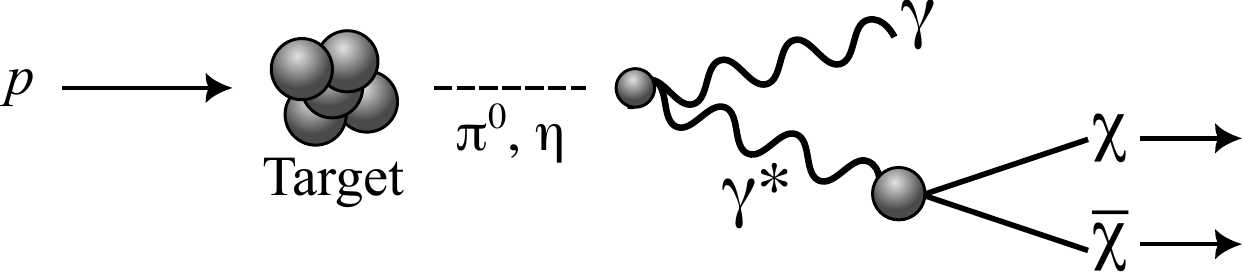}
    \caption{Schematic of millicharged particle production in the NuMI beam via a meson decay and a virtual photon, figure modified from \cite{schematic_2017_mcp}.}
    \label{fig:schematic_decay}
\end{figure}

\begin{figure}[t]
    \centering
    \includegraphics[width=0.46\textwidth]{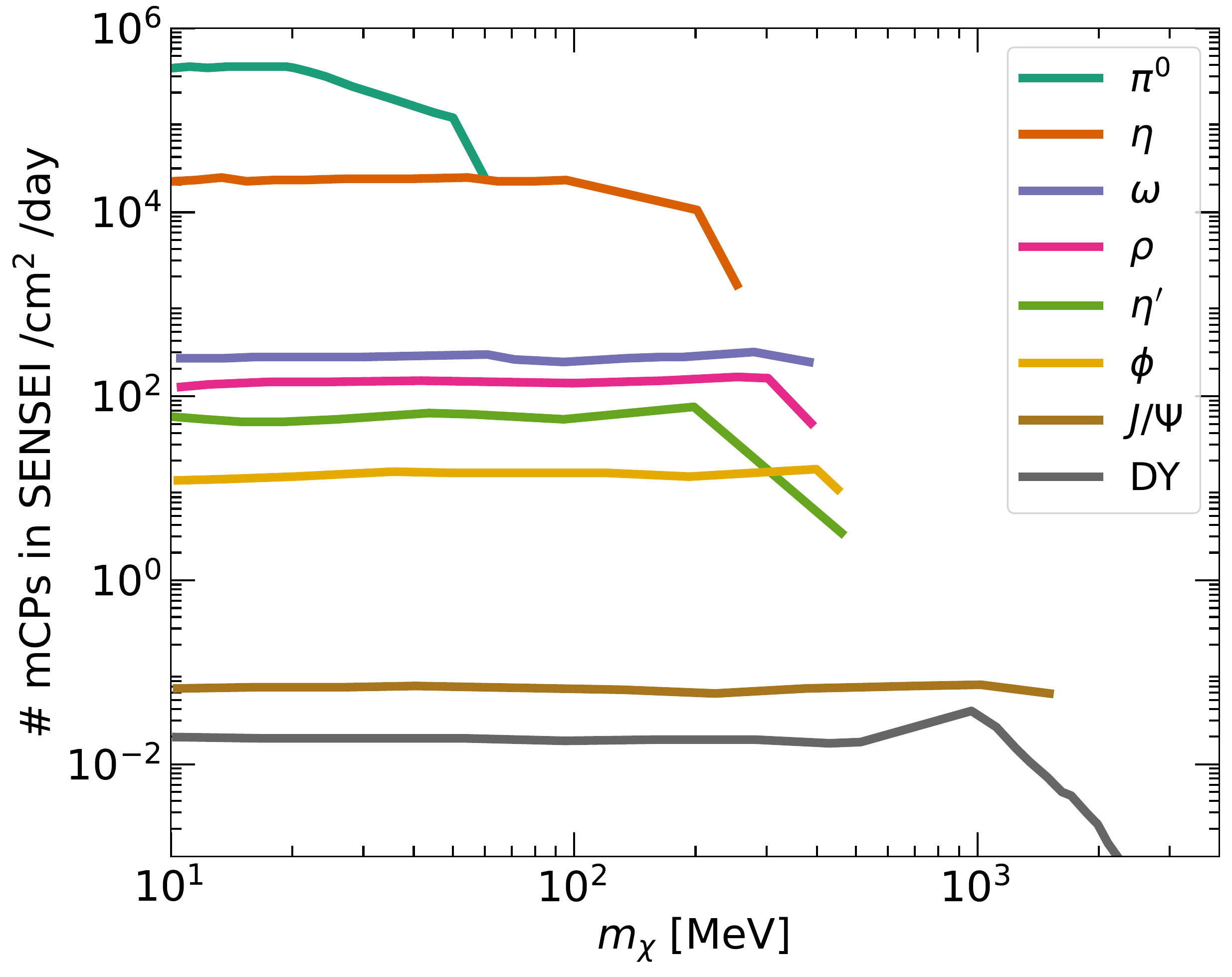}
    \caption{Number of mCPs accepted by the SENSEI@MINOS detector (9.216 cm $\times$ 1.329 cm) placed 1\,km away from the target integrated over all mCP energies for each production channel. This flux calculated assuming $\epsilon=10^{-2}$. Adapted from~\cite{mcpRoni2019}. 
    }
    \label{fig:accepted}
    \end{figure}

\noindent In a wide range of neutrino experiments, accelerator-based neutrino beams are sourced by a high-intensity proton beam colliding with a fixed target. If mCPs exist, they may be produced collinearly with the neutrino beam in photon-mediated decays of scalar mesons, vector mesons, and direct Drell-Yan processes resulting from each collision (Fig.~\ref{fig:schematic_decay}). The NuMI beamline provides 120~GeV protons, which strike a fixed, graphite target. Following Ref.~\cite{mcpRoni2019}, where mCP production is calculated using Pythia8, we show in Fig.~\ref{fig:accepted} the flux of mCPs created in the NuMI beamline for each decay product, integrated over the entire energy range that would reach the SENSEI detector. As the mass of the mCP increases, decays into millicharged pairs become kinematically inaccessible, and the flux reaching the detector is greatly reduced.  The number of protons on target (POT) used for this calculation was taken from beam operation data at Fermilab during the SENSEI data-taking period from Feb. 25, 2020 to March 19, 2020 and amounted to an average of $2.4\times10^{18}$ per day. The number of POT per day fluctuated by less than 20\%, while the image-by-image efficiency fluctuated by less than 10\%. Therefore, we will calculate the limits using the mean values of the number of POT per day and the image-by-image efficiency, which is a good approximation of the total POT expected during the experiment.

Of the 1~km between the NuMI target and the SENSEI detector, approximately 500~m is dirt, which the mCPs must traverse before reaching the detector. As is discussed in Ref.~\cite{mcpRoni2019}, this leads to two effects that must be considered: energy losses to the material and angular deflections caused by the interactions. To treat these two effects, it is important to distinguish between two types of interactions between mCPs and matter: soft collisions, which dominate the number of low energy scatters experienced by the particle, and the rare hard collisions, which dominate energy losses. Soft collisions will be discussed in detail in the following sections. Due to the energy ($\sim$GeV) of the mCPs produced by the beam and the probability of interaction given by the Bethe-Fano formula for high energy recoils, it is expected that the particles would lose only about an order of $\sim$MeV of energy, which would not alter the energy spectrum of the mCPs nor the number of particles reaching the detector by a significant amount. The angular deflection is also small enough to be negligible for our analysis. For mCPs deflected due to soft collisions with nuclei, the maximum possible scattering is of order $\varepsilon$~\cite{mcpRoni2019}.
mCPs produced by protons in the NuMI beamline would be highly boosted, and given that the SENSEI detector has a relatively small area of $A$=12.25\,cm$^2$, they will have a spatially uniform flux over the entire detector. 

\section{Interaction of \MakeLowercase{m}CPs with silicon  \label{sec:detection}}
The theory of electromagnetic excitations from passing point-like charged particles has a long history dating back to early work by Bethe, Fermi, and Landau, among others~\citep[][and references therein]{Bethe:1930ku,FermiEnergyLoss40,Landau:1944if, Vavilov:1957zz, AllisonCobb80}. Much of the modern literature focuses on the energy loss per unit length, or ${\rm d} E/{\rm d} x$, which is typically dominated by ionization (for muons or non-relativistic electrons) or bremsstrahlung (for relativistic electrons). This is the relevant quantity when describing the statistical properties of energy loss by particles with ``large'' electric charges. For our current purposes, we are interested in the limit in which only countably few interactions take place. One must therefore consider instead the process that produces the largest \textit{event rate} that is available with the SENSEI data. Crucially, this is distinct from the dominant energy loss mechanism. As we outline below, this necessitates a proper treatment of collective modes, and in particular silicon's bulk plasmon excitation. 
A more detailed discussion of low-threshold detection of highly relativistic particles with silicon detectors will appear elsewhere~\cite{Plasmon:forthcoming}, and we provide here only the salient features. 

An appropriate theory of energy losses for a passing relativistic charged particle was first provided by Fermi~\cite{FermiEnergyLoss40}. This theory treats the passing mCP as a classical source of electromagnetic fields and incorporates the relevant bulk material properties via a complex dielectric function. The formalism is exact in the eikonal limit, defined by $k_\mu \ll p_\mu$, where $k_\mu$ is the four-momentum transfer to the target and $p_\mu$ is the mCP four-momentum, and where the particle can be approximated by a straight-line trajectory (see e.g.\ \S 13.6 of~\cite{Weinberg:1995mt}).  The effective interaction cross section between mCPs and silicon is given by~\cite{FermiEnergyLoss40} ({\it c.f.} Eq.\ (15) of~\cite{AllisonCobb80}) 
\begin{equation}    
    \label{energy-loss-fermi}
    \begin{split}
     \dv{\sigma}{\omega} = &\frac{8\alpha\varepsilon^2}{n \beta^2}\int_0^\infty dk \bigg\{\frac{1}{k}\mathrm{Im}\left(-\frac{1}{\epsilon(\omega, k)}\right)\\
     & ~+k\left(\beta^2 - \frac{\omega^2}{k^2}\right)\mathrm{Im}\left(\frac{1}{-k^2 + \epsilon(\omega, k)\omega^2}\right)\bigg\}~, 
   \end{split}
\end{equation}
where $\varepsilon$ is the mCP's charge, $\alpha$ the fine structure constant, $\beta=p/E$ the velocity of the mCP, $k_\mu = (\omega, \vb{k})$, $n$ the number of electrons per unit volume, and $k=|\vb{k}|$ in the integration above. Relativistic effects related to e.g.\ the field contraction of a highly boosted particle are treated exactly, and we have retained both transverse and longitudinal components of the field. Non-relativistic treatments typically drop the transverse contribution and retain only the term proportional to the energy loss function (ELF), ${\rm Im}(-1/\epsilon)$, where $\epsilon(\omega,k)$ is the dielectric function of the material. The bulk plasmon excitation of silicon manifests itself as a zero in the real part of the dielectric function. 

We have used tabulated GPAW calculations of the dielectric function for silicon, as can be obtained from the DarkELF repository~\cite{KnapenDarkELF21} to compute the cross section from Eq.\ \eqref{energy-loss-fermi}. These calculations were validated against publicly available electron energy loss spectroscopy (EELS) data, resulting in a good level of agreement~\cite{MichaelKEELS, SiEELS}. Other approaches, such as the photoabsorption ionization (PAI) model~\cite{AllisonCobb80}, which relies on free-particle approximations and photo-absorption data, were found to underestimate the cross section by as much as $\sim$40\% for the energy range relevant to our analysis ($\sim$1.2--20~eV). As alluded to above, the most important feature in the silicon dielectric function is the bulk plasmon~\cite{MichaelKEELS, SiEELS}. This is easily understood, since the response of the system is resonantly enhanced, and provides electronic excitations above the SENSEI threshold in a region where background events are observed. Unlike in DM direct detection, where plasmons have received substantial attention~\cite{Kurinsky:2020dpb,Kozaczuk:2020uzb,Hochberg:2021pkt,Boyd:2022tcn}, the kinematics of boosted particles from accelerator beams are such that the plasmon is easily accessible kinematically. 
Setting $\beta=1$ in Eq.\ \eqref{energy-loss-fermi}, we can write $\sigma_{\rm int}$ as the interaction cross section defined via, 
\begin{equation}
    \sigma_{\rm int}  = \int_{\omega_0}^\infty \dd \omega \frac{{\rm d} \sigma}{{\rm d}  \omega} (\beta=1)~.
    \label{Cross_int}
\end{equation}

\section{\MakeLowercase{m}CP detection with SENSEI} \label{sec:limit}
Once the mCP interacts with the detector, the likelihood of an electron recoil ionizing $1-6~e^-$ in the detector, which corresponds to ($\sim$1.2--20~eV) energy depositions, is determined by the model outlined in~\cite{Ramanathan2020}. In order to include this effect that directly impacts energy reconstruction, the cross-section calculated via Eq.~\eqref{Cross_int} must be convolved with the probability of producing electron-hole pairs for each individual channel yielding the detection cross section $\sigma_{\rm det}$.

We calculate the expected number of events at our detector as
\begin{equation}
    N(\varepsilon,m_{\chi})=A  \Delta T \int \phi(\varepsilon,E_{\chi} ,m_{\chi}) P(hits \geq 1) dE_{\chi},
    \label{Number_ev}
\end{equation}

\noindent where $A$ is the detector area, $\phi(\varepsilon,E_{\chi},m_{\chi})$ is the flux of millicharged particles, $\Delta T$ is the total exposure and 
\begin{equation}
    P(hits \geq 1)=1-e^{-L/\lambda},
\end{equation}
is the probability of interaction set by the mean free path $\lambda=(n_e\sigma_{\rm det})^{-1}$, which depends on the particle millicharge $\varepsilon$ and the electron number density $n_e$. 

When $L/\lambda$ is small, $P(hits \geq 1)\approx L/\lambda$. Substituting this probability in Eq.~\eqref{Number_ev} makes it clear that the number of expected events depends solely on the detector volume, and that no geometric effects need to be considered. For a general discussion involving geometric effects, see Ref~\cite{Perez:2023awk}.

In addition, since the bulk of the flux from an accelerator source is highly boosted, to a very good approximation, Eq.~\eqref{Number_ev} can be reduced to

\begin{equation}
    N(\varepsilon,m_{\chi}) =  N_e \Phi_{\rm fast}(\varepsilon,m_{\chi}) \sigma_{\rm det}~\label{rate-simple}, 
\end{equation}
where $\Phi_{\rm fast}(\varepsilon,m_{\chi})$ is defined as the total flux of mCPs with boosts larger than $\gamma=3$ (as shown in Fig.~\ref{fig:accepted}), and $N_e$ is the number of electrons in the SENSEI detector. The reduced form of Eq.~\eqref{rate-simple} underscores the fact that the limits set by SENSEI are primarily sensitive to the integrated flux of mCPs, and are insensitive to the detailed shape of the spectrum.

\begin{figure}[H]
    \hspace{-3mm}
    \includegraphics[width=0.485\textwidth]{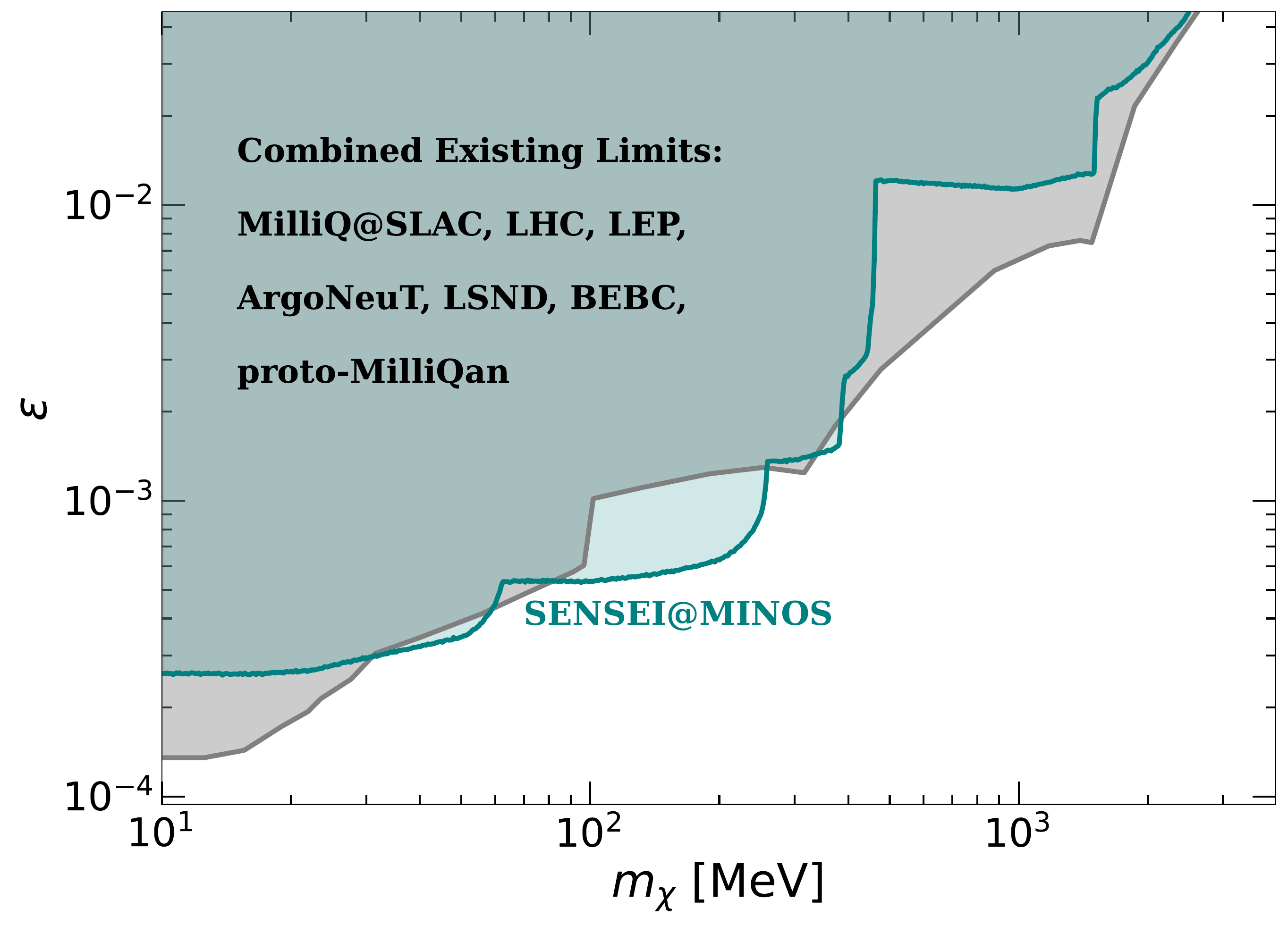}
    \caption{ Cyan line/region shows the 95\%~C.L.~limit on mCPs from the SENSEI data collected in the MINOS cavern in 2020. Gray line/region shows constraints from other experiments \cite{MilliQSLAC,Argoneut_limit,LEP,Magill:2018tbb,ball2020search}}
    \label{fig:lim}
\end{figure}

In Fig.~\ref{fig:lim}, we show the 95\% C.L.~constraints on the mCP parameter space from the published SENSEI data, which is reproduced in Table~\ref{tab:senseiEventCount}, and compare these constraints with the existing bounds from other experiments. This limit was calculated for each electron channel independently, taking into account their different backgrounds and then combined using a frequentist approach based on the likelihood ratio as in~\cite{Barak:2020fql}. We see that we improve on previous bounds by as much as a factor of 2 in the mass range 100~MeV to 210~MeV.

\section{Conclusion} \label{sec:Conclusion}

In this work, we set new constraints on millicharged particles using data from the SENSEI 2020 run~\cite{Barak:2020fql}. The very low background and low detection threshold of the Skipper-CCD installed in the MINOS cavern at Fermilab enables a very sensitive search for these proposed particles.

By utilizing a validated GPAW calculation against publicly available electron energy loss spectroscopy data, the interaction cross-section between mCPs and silicon was computed. The bulk plasmon effects play a critical role in this interaction due to the fact that the particles coming from the NuMI beamline can easily access the plasmon energies and excite electrons well above the SENSEI threshold.

This result provides the most stringent limits to date in the mass range from 30~MeV to 380~MeV, and establishes Skipper-CCDs as a very promising technology to probe new parameter space for millicharged particles in future experiments.
\\

\section*{Acknowledgements}

\noindent We are grateful for the support of the Heising-Simons Foundation under Grant No.~79921.
This work was supported by Fermilab under U.S.~Department of Energy (DOE) Contract No.~DE-AC02-07CH11359. 
The CCD development work was supported in part by the Director, Office of Science, of the DOE under No.~DE-AC02-05CH11231. RE acknowledges support from DOE Grant DE-SC0009854 and Simons Investigator in Physics Award~623940. 
The work of TV and EE is supported by the I-CORE Program of the Planning Budgeting Committee and the Israel Science Foundation (grant No.1937/12). TV is further supported  by the European Research Council (ERC) under the EU Horizon 2020 Programme (ERC- CoG-2015 -Proposal n.~682676 LDMThExp), and a grant from The Ambrose Monell Foundation, given by the Institute for Advanced Study.
The work of SU is supported in part by the Zuckerman STEM Leadership Program.
IB is grateful for the support of the Alexander Zaks Scholarship, The Buchmann Scholarship, and the Azrieli Foundation. RP is funded by the Neutrino Theory Network Program Grant under Award Number DEAC02-07CHI11359 and the US DOE under Award Number DE-SC0020250. RP is also supported in part by the U.S. Department of Energy, Office of Science, Office of High Energy Physics, under Award Number DE-SC0011632 and by the Walter Burke Institute for Theoretical Physics. This manuscript has been authored by Fermi Research Alliance, LLC under Contract No. DE-AC02-07CH11359 with the U.S.~Department of Energy, Office of Science, Office of High Energy Physics. The United States Government retains and the publisher, by accepting the article for publication, acknowledges that the United States Government retains a non-exclusive, paid-up, irrevocable, world-wide license to publish or reproduce the published form of this manuscript, or allow others to do so, for United States Government purposes. The work of Zhen Liu was supported in part by the DOE grant DE-SC0022345.

\bibliography{mCPsensei.bib}

\end{document}